\documentclass[12pt]{article}
\usepackage{amsmath,amssymb,epsfig}
\begin{document}
\parindent=0pt
\parskip=6pt
\rm

\vspace*{0.3cm}

\begin{center}
{\bf NEW FEATURES OF THE PHASE TRANSITION TO SUPERCONDUCTING STATE IN
THIN FILMS}

\vspace{0.5cm}

 D. V. SHOPOVA$^{\ast}$, T. P. TODOROV$^{\dag}$

{\em  CPCM Laboratory, Institute of Solid State Physics,\\
 Bulgarian Academy of Sciences, BG-1784 Sofia, Bulgaria.} \\
\end{center}

$^{\ast}$ Corresponding author: sho@issp.bas.bg

$^{\dag}$ Permanent address: Joint Technical College at the Technical
University of Sofia.

\vspace{0.7cm}

{\bf Key words}: fluctuations, latent heat, order parameter, equation
of state.

{\bf PACS}: 05.70.Jk, 74.20.De, 74.78.Db,

\vspace{0.2cm}

\begin{abstract}
The Halperin-Lubensky-Ma (HLM) effect of a fluctuation-induced change
of the order of phase transition in thin films of type I
superconductors with relatively small Ginzburg-Landau number $\kappa$
is considered. Numerical data for the free energy, the order parameter
jump, the latent heat, and the specific heat of W, Al and In are
presented to reveal the influence of  film thickness and  material
parameters on the properties of the phase transition. We demonstrate
for the first time that in contrast to the usual notion the HLM effect
occurs in the most distinct way in superconducting films with high
critical magnetic field $H_{c0}$  rather than in materials with small
$\kappa$. The possibility for an experimental observation of the
fluctuation change of the order of superconducting phase transition in
superconducting films is discussed.
\end{abstract}

{\bf 1. Introduction}

Our paper is intended to clarify the best conditions for an
experimental observation of the Halperin-Lubensky-Ma effect
(HLM)~\cite{Halperin:1974, Chen:1978} of a fluctuation-induced
first-order phase transition from normal to Meissner phase in a zero
external magnetic field for thin films of type I
superconductors~\cite{Folk:2001,Shopova:2002,Shopova1:2003} . For this
purpose  we present new theoretical results about the thermodynamics of
the phase transition from normal to superconducting state in a  zero
external magnetic field.

The HLM effect is predicted  theoretically  ~\cite{Halperin:1974} to
occur  in pure~\cite{Chen:1978, Shopova:2002,Lawrie:1982} and
disordered~\cite{Uzunov:1983,Athorne:1985,Busiello:1986} bulk,
(three-dimensional - 3D), and 2D~\cite{Lovesey:1980} superconductors,
as well as in quasi-2D superconducting
films~\cite{Folk:2001,Shopova1:2003} but up to now it has not been
observed in experiments. The calculated effect is very small in 3D
superconductors and is not possible to be detected  even for a high
purity of the sample and perfection of the crystal
lattice~\cite{Halperin:1974,Chen:1978,Shopova:2002}. Recently, it has
been shown~\cite{Folk:2001,Shopova1:2003} that in thin (quasi-2D) films
the HLM effect is much stronger than in 3D samples and could be
observed by available experimental techniques if  the type of
superconductor and film thickness are properly chosen for the
experiments; for a review, see also, Refs.~\cite{Uzunov1:1993,
Folk1:1999}. This result gives an opportunity to search for the effect
in suitable superconducting films.

The HLM effect appears as  a result of the interaction between the
superconducting order parameter $\psi(\vec{x})$ and the vector
potential $\vec{A}(\vec{x})$ of the magnetic induction in the
Ginzburg-Landau (GL) free energy of a superconductor. According to the
theoretical paradigm introduced for the first time in the scalar
electrodynamics by Coleman and Weinberg (CW)~\cite{Coleman:1973}, this
effect should occur in all physical systems described by Abelian-Higgs
models where a scalar gauge field (like $\psi$ in superconductors)
interacts in a gauge invariant way with another vector gauge field
(like the vector potential $\vec{A}$ in superconductors). In addition
to the mentioned examples of superconductors and scalar
electrodynamics, the same type of interaction plays an important  role
in the nematic-smectic A phase transition in liquid
crystals~\cite{Lubensky:1978,Anisimov:1990,Yethiraj:2000} and phase
transitions in the early universe~\cite{Linde:1979}. HLM effect may be
 also relevant to quantum phase transitions in
superconductors~\cite{Fisher:1988,Fisher1:1990,Shopova2:2003} and
itinerant ferromagnets~\cite{Kirkpatrick:2003}.

On the other side, there are certain theoretical investigations, based
on Monte Carlo simulations~\cite{Bartholomew:1983} and the so-called
``dual model"~\cite{Kiometzis:1995} which do not confirm the
fluctuation-change of the order of the phase transition (see also
Refs.~\cite{Folk1:1999}). That is why, extensive experiments intended
to verify the existence of the effect were made in liquid crystals;
see, e.g., Refs.~\cite{Folk1:1999}. But in liquid crystals the
weakly-first order phase transition predicted by CW and HLM can be
obscured by similar effects due to the strong crystal anisotropy, while
the recent result~\cite{Folk:2001} about the considerable enhancement
of HLM effect in suitable superconducting films can be used for more
reliable experiments. For this aim we need to find the best material
parameters and the most suitable film thickness, having in mind some
purely experimental problems that may appear.

Recently,  we partly solved the problem for Al
films~\cite{Shopova1:2003} nevertheless additional theoretical
investigations should be done. In this paper we shall present some new
theoretical predictions for Al films as well as new numerical data for
thin films of W and In. The choice of this element superconductors is
made for their relatively small GL number $\kappa = (\lambda/\xi)$,
which allows  a more distinct appearance of the HLM effect in both bulk
and thin film superconductors~\cite{Halperin:1974,
Folk:2001,Shopova1:2003}; here $\lambda$ is the London penetration
depth and $\xi$ is the coherence length~\cite{Lifshitz:1980}.

 We focus our attention on numerical data
for the behavior of the free energy and directly measurable
thermodynamic quantities like the order parameter jump, the latent
heat, and the specific heat. A surprising result of our analysis of the
data for W, Al, and In is that the HLM effect in thin films is stronger
in  case of relatively high zero-temperature critical magnetic field
$H_{c0}$ rather than for relatively small GL number $\kappa$, as
claimed in preceding papers~\cite{Halperin:1974, Chen:1978, Folk:2001,
Shopova:2002,Shopova1:2003}.

Our investigation is based on  the theoretical results from preceding
papers~\cite{Halperin:1974, Folk:2001,Shopova:2002, Shopova1:2003}. In
Sec.~2 we shall outline the theoretical framework of our study. In
Sec.~3 we present our analysis of  thin films of tungsten (W),
aluminium (Al), and indium (In) and a discussion of the results with a
special emphasis on their application to experiments. In Sec.~4 we
summarize our findings.

{\bf 2. Theoretical basis}

Our investigation is based on the Ginzburg-Landau free
energy~\cite{Lifshitz:1980} of a D-dimensional superconductor with
volume $V = (L_1...L_D)$ given by
\begin{equation}
\label{eq1}
 F = \int d^D x \left[ a|\psi|^2 + \frac{b}{2}|\psi|^4 +
\frac{\hbar^2}{4m}\left|\left(\nabla - \frac{2ie}{\hbar
c}\vec{A}\right)\psi \right|^2 + \frac{1}{16\pi}\sum_{i,j=1}^{3} \left(
\frac{\partial A_i}{\partial x_j} - \frac{\partial A_j}{\partial x_i}
\right)^2 \right]\;,
\end{equation}
where $a = \alpha_0(T-T_{c0})$ and $b > 0$, are the Landau parameters,
$e=|e|$ is the electron charge, $\psi(\vec{x})$ is the order parameter,
and $\vec{A}(\vec{x})$ is the vector potential of the magnetic field.

The critical temperature $T_{c0}$ corresponds to the second order phase
transition  which occurs in a  zero external magnetic field when the
fluctuations, $\delta\varphi(\vec{x})$ and $\delta\vec{A}(\vec{x})$, of
both fields $\psi(\vec{x})$ and $\vec{A}(\vec{x})$ are neglected.
Usually, this case  is considered in the low-temperature
superconductors, where the effect of the superconducting fluctuations
$\delta \varphi(\vec{x})$ on the thermodynamics is very small and
practically uninteresting, and the same has been supposed for the
magnetic fluctuations $\delta\vec{A}(\vec{x})$ before the appearance of
HLM paper~\cite{Halperin:1974}. In our study the superconducting
fluctuations are ignored as negligibly small which is a suitable
approximation in type I superconductors where $\lambda \ll \xi$. But we
take into account the magnetic fluctuations to a full extent. Then the
normal-to-superconducting phase transition in a zero external magnetic
field turns out of first order at an equilibrium phase transition
temperature $T_{\mbox{\scriptsize eq}}$ that is different from
$T_{c0}$. So, our task will be to point the type of superconductors,
where this picture may be valid and investigate the properties of the
first order phase transition.

We shall follow the theoretical approach described in details in
preceding papers~\cite{Halperin:1974,
Chen:1978,Folk:2001,Shopova1:2003}, where an effective free energy
$F_{\mbox{\footnotesize eff}}(\psi)$ of the type I superconducting film
was obtained~\cite{Shopova1:2003}. There we  also neglected  the
superconducting fluctuations and did the calculation of the mean-field
value of the uniform ($\vec{x}$-independent) superconducting order
parameter $\psi$  in a self-consistent way after taking into account
the magnetic fluctuations through an exact integration out of the field
$\vec{A}(\vec{x})$ in the partition function of the superconductor. As
the external magnetic field is equal to zero the regular part
$\vec{A}_0 = (\vec{A} - \delta \vec{A})$ of $\vec{A}$ related to it can
be set equal to zero, too. Then $\vec{A} = \delta\vec{A}$, therefore,
we consider the net effect of the magnetic fluctuations.

 Having in mind
that in type I superconductors a stable vortex phase cannot occur, we
again can assume that the order parameter $\psi$ that describes the
uniform Meissner phase in the bulk of the superconducting film is
$\vec{x}$-independent.  Our investigation is based on the
quasi-macroscopic GL theory so we must consider films of thickness $
L_0 \gg a_0 $, where $a_0$ is the lattice constant. In such films the
surface energy can be ignored and one can use periodic boundary
conditions without a substantial departure from the real situation. In
this way, the surface effects as a source of a spatial dependence of
the order parameter $\psi$ are also eliminated.

Following~\cite{Halperin:1974,Chen:1978,Folk:2001,Shopova:2002,
Shopova1:2003} we present the effective free energy density $f(\psi) =
F_{\mbox{\footnotesize eff}}(\psi) /V$ of a 3D superconducting slab of
volume $V = (L_1L_2L_0)$ and thickness $L_0$ in the form:
\begin{equation}
\label{eq2}
 f(\varphi) =\frac{H^2_{c0}}{8\pi}\left\{2t_0\varphi^2 +
\varphi^4 + C(1 +
t_0)\left[\left(1+\mu\varphi^2\right)\mbox{ln}\left(1+\mu\varphi^2\right)
- \mu\varphi^2\mbox{ln}\left(\mu\varphi^2\right)\right]\right\}\:,
\end{equation}
where
\begin{equation}
\label{eq3}
 C= \frac{2\pi^2k_BT_{c0}}{L_0\xi_o^2H^2_{c0}}\:.
\end{equation}
Here $\varphi = (|\psi|/|\psi_0|)$ is the dimensionless order parameter
defined with the help of the zero-temperature value $|\psi_0| = |\psi(T
= 0)| = (\alpha_0T_{c0}/b)^{1/2}$ of $|\psi|$,  $t_0 =
(T-T_{c0})/T_{c0}$, $\mu = (\xi_0/\pi\lambda_0)^2$ is given by the
zero-temperature value $\xi_0 = (\hbar^2/4m\alpha_0T_{c0})^{1/2}$ of
$\xi$ and  $\lambda_0 = (b/\rho_0\alpha_0T_{c0})^{1/2}$ is the
zero-temperature penetration depth; ($\rho_0 = 8\pi e^2/mc^2$). We also
use the notations: $\lambda (T) = \lambda_0/|t_0|^{1/2}$ and $\xi (T) =
\xi/|t_0|^{1/2}$. The critical magnetic field at $ T= 0$ is given
by~\cite{Lifshitz:1980} $H_{c0} = \alpha_0T_{c0}(4\pi/b)^{1/2}$. The
relations of $H_{c0}$ and $\xi_0$ with $b$ and $\alpha_0$,
respectively, can be used together with the experimental data for
$H_{c0}$ and $\xi_0$ in concrete superconducting substances in order to
calculate the parameters $b$ and $\alpha_0$.

The equilibrium order parameter $\varphi_0 > 0$ corresponding to the
Meissner phase can be easily obtained from the equation $\partial
f(\varphi)/\partial\varphi = 0$ and Eq.~(\ref{eq2}):
\begin{equation}
\label{eq4}
 t_0 + \varphi^2_0 + \frac{C\mu(1+t_0)}{2}\left[\mbox{ln}\left(1 +
 \frac{1}{\mu\varphi_0^2}\right)
- \frac{1}{1+\mu\varphi_0^2}\right] = 0\:.
\end{equation}

 The logarithmic divergence in Eq.~(\ref{eq4}) has no chance to occur because $\varphi_0$ is
  always positive and does not tend to zero.

 We shall use the notations from Ref.~\cite{Shopova1:2003} for the
 entropy jump $\delta s$ and the specific
heat jump $\delta C$ at the equilibrium phase transition point
$T_{\mbox{\scriptsize eq}}$ of first order corresponding to a zero
external magnetic field.   Here we shall give the previously calculated
results for the leading terms in these quantities (terms of higher
order are neglected as small), namely,
\begin{equation}
\label{eq5}
 \delta s = - \frac{H_{c0}^2}{4\pi T_{c0}} \varphi^2_{\mbox{\scriptsize
eq}}\:,
\end{equation}
and
\begin{equation}
\label{eq6}
 \delta C = \frac{H^2_{c0}}{4\pi T_{c0}} \:.
\end{equation}
The latent heat of the phase transition is given by $Q =
T_{\mbox{\scriptsize eq}}\delta s$ and Eq.~(\ref{eq4}). Since
 the temperatures $T_{\mbox{\scriptsize eq}}$ and $T_{c0}$ have
very close values, the difference between the values of $Q$, $\delta
s$, and $\delta C$ at $T_{c0}$ and $T_{eq}$, respectively, can also be
ignored, for example, $|\delta C(T_{\mbox{\scriptsize eq}})-\delta
C(T_{c0})|/\delta C(T_{c0}) \ll 1$ and we can use either $\delta
C(T_{c0})$ or $\delta C(T_{\mbox{\scriptsize eq}})$
~\cite{Shopova1:2003}. Here the jumps $\delta s$, $\delta C$, and $Q$
are all taken at the equilibrium phase transition value
$T_{\mbox{\scriptsize eq}}$ but we shall not supply them with the
subscript ``eq" as we do for other quantities.

 Eqs.~(\ref{eq2})~-~(\ref{eq6}) are valid for thin films ($a_0 \ll
 L_0 \sim \xi_0$) in a  zero external magnetic field $\vec{H}$ and for negligibly
small $ \psi $-fluctuations which means that they are applicable for
low-temperature nonmagnetic superconductors ($T_{c0} < 20 K$). Because
in experiments the external magnetic field cannot be completely
eliminated, vortex states may occur for $H= |\vec{H}| > 0$  below $T_c=
T_c(H) \leq T_{c0}$ in type II superconducting films and this will
obscure the HLM effect. Note, that the magnetic field $H$ generates
additional entropy jump at the phase transition point $T_c(H)$ and this
effect can hardly be separated from the entropy jump~(\ref{eq5}) caused
by the magnetic fluctuations in the close vicinity of $T_{c0}$.
Therefore, in experiments intended to a search of the HLM effect we
must choose type I superconductors. The second important point is
connected with the value of the square $\varphi^2_{\mbox{\scriptsize
eq}}$ which is proportional to the superconducting current ($j_s \sim
|\psi|^2_{\mbox{\scriptsize eq}} \sim \varphi^2_{\mbox{\scriptsize
eq}}$\cite{Lifshitz:1980}) and to the equilibrium jumps $\delta s$ and
$\delta Q$. In 3D superconductors the ratio ($Q/\delta C)$ depends on
$H^2_{c0}\xi_0 \sim \epsilon_c\kappa^{-6}$, where $\epsilon_c \sim
10^{-16}$ denotes the extremely small Ginzburg-Levanyuk critical
region~\cite{Uzunov1:1993} of low-temperature superconductors
~\cite{Halperin:1974}. Therefore, the latent heat in these 3D
superconductors can hardly be observed in experiments. But in thin
films the substantial dependence of the entropy $\delta s$ and the
specific heat $\delta C$ is  on the critical magnetic field $H^2_{c0}$,
as shown by Eqs. (\ref{eq5})~-~(\ref{eq6}) and the analysis in Sec.~3.

The equations~(\ref{eq2}) and (\ref{eq4}) corresponding to quasi-2D
films are quite different from the respective
equations~\cite{Halperin:1974,Chen:1978} for bulk (3D-) superconductors
but it is easily seen that the relatively large values of the order
parameter jump $\varphi^2_{\mbox{\scriptsize}}$ in thin films again
correspond to relatively small values of the GL parameter $\kappa$.
That is why we consider element superconductors with small values of
$\kappa$  and study the effect of this parameter, the critical magnetic
field $H_{c0}$ and the film thickness $L_0$ on the properties of the
fluctuation-induced first order phase transition.

Theoretical results  we have used in this Section for quasi-2D
superconducting films  are consistent with the theory~\cite{Craco:1999}
of 2D-3D crossover phenomena near phase transition points and the 2D-3D
crossover theory~\cite{Rahola:2001,Shopova3:2003} of the HLM effect;
see also Ref.~\cite{Abreu:2003}.

{\bf 3. Results and discussion}

We use experimental data for $T_{c0}$, $H_{c0}$, $\xi_0$ and $\kappa$
for W, Al, and In published in Ref.~\cite{Madelung:1990} (see Table 1).
In some cases the GL parameter $\kappa$ can be calculated with the help
of the relation  $\kappa = (\lambda_0/\xi_0)$ and the available data
for $\xi_0$ and $\lambda_0$. In other cases it is more convenient to
use the following representation of the zero-temperature penetration
depth:
\begin{equation}
\label{eq7}
  \lambda_0 = \frac{\hbar c}{2\sqrt{2}eH_{c0}\xi_0}\:.
\end{equation}
The value of $|\psi_0|$ in Table~1 is found from
 \begin{equation}
\label{eq8}
  |\psi_0| = \left(\frac{m}{\pi \hbar^2}\right)^{1/2}\xi_0H_{c0}\:.
\end{equation}
Eqs.~(\ref{eq7}) and (\ref{eq8}) are obtained from the formulae given
after Eq.~(\ref{eq3}). Besides we calculate the parameter $\tilde{C} =
CL_0$ with the help of Eq.~(\ref{eq3}) and the data in Table~1.

 Note, that the experimental
data vary within 5-10\% depending on the experimental technique used in
measurements . Moreover, these data correspond to bulk samples and may
differ within 10-20\% from those for very thin films ($L_0 <
10^{-2}\mu$m). However, these variations in the experimental data do
not essentially affect our results.

\vspace{0.5cm}
 \small
 Table 1. Values of $T_{c0}$, $H_{c0}$, $\xi_0$,
$\kappa$, and $|\psi_0|$ for W, Al, In.\\
\begin{tabular}{cccccc}
\hline substance & $T_{c0}$ (K) & $H_{c0}$ (Oe)& $\xi_0$ ($\mu$m) &
$\kappa$ & $|\psi_0|\times 10^{-11}$  \\ \hline W & $0.015$ & $1.15$
& $37$ & $0.001 $& $0.69$  \\
\hline Al & $1.19$ & $99.00$ & 1.16 & $0.010$ &
$2.55$ \\
\hline In & 3.40 & $281.5$ & 0.44 & 0.145 & $2.0$
\\
\hline
\end{tabular}

\vspace{0.5cm}

\small Table 2. Values of $t_{\mbox{\scriptsize eq}}$,
$\varphi_{\mbox{\scriptsize eq}}$,
  and $Q$ (erg/cm$^3$) for films of W, Al, and In
   with different\\ thicknesses $L_0$~($\mu$m). \\
\small
\begin{tabular}{lccccccccc}
\hline
&\multicolumn{3}{c}{Al}&\multicolumn{3}{c}{In}&\multicolumn{3}{c}{W}
\\\cline{2-10}
 $L_0$ &$\mbox{t}_{\mbox{\scriptsize eq}}$
& $\mbox{$\varphi$}_{\mbox{\scriptsize eq}}$ & $Q$
&$\mbox{t}_{\mbox{\scriptsize eq}}$ & $\varphi_{\mbox{\scriptsize eq}}$
& $Q$ &$\mbox{t}_{\mbox{\scriptsize eq}}$ &
$\varphi_{\mbox{\scriptsize eq}}$ & $Q$ \\
\hline
 0.05 &$-0.00230$ & $0.041$ & $-1.95$ &$-0.00167$ &
$0.025$ & $-3.94$ &$-0.00174$ & $0.039$ & $-1.6\times 10^{-4}$
\\\hline 0.1 &$-0.00147$ & $0.032$ & $-0.80$
&$-0.00094$ & $0.017$ & $-1.82$ &$-0.00118$ & $0.032$ & $-1.1\times
10^{-4}$
\\\hline 0.3&$-0.00070$ & $0.023$ & $-0.41$
&$-0.00037$ & $0.010$ & $-0.63$ &$-0.00064$ & $0.023$ & $--5.6\times
10^{-5}$
\\\hline 0.5 &$-0.00048$ & $0.016$ & $-0.20$
&$-0.00029$ & $0.008$ & $-0.40$ &$-0.00048$ & $0.020$ & $-4.1\times
10^{-5}$
\\\hline 1 &$-0.00029$ & $0.012$ & $-0.11$
&$-0.00013$ & $0.006$ & $-0.23$ &$-0.00032$ & $0.016$ & $-2.7\times
10^{-5}$
\\\hline 2 &$-0.00017$ & $0.009$ & $-0.06$
&$-0.00008$ & $0.004$ & $-0.10$ &$-0.00021$ & $0.013$ & $-1.8\times
10^{-5}$ \\
\hline
\end{tabular}

\vspace{0.3cm} \normalsize

The order parameter profile for Al films of different thicknesses is
shown in Fig.~1. It is readily seen that the behavior of the function
$\varphi_0(t_0)$ corresponds to a well established phase transition of
first order. The vertical dashed lines in Fig.~1 indicate the
respective values of $t_{\mbox{\scriptsize eq}} =
t_0(T_{\mbox{\scriptsize eq}})$, at which the equilibrium phase
transition occurs as well as the equilibrium jump
$\varphi_0(T_{\mbox{\scriptsize eq}})= \varphi_{\mbox{\scriptsize eq}}$
for different thicknesses of the film. The parts of the
$\varphi_0(t_0)$-curves which extend up to $t_0 > t_{\mbox{\scriptsize
eq}}$ describe the metastable (overheated) Meissner states which can
appear
 under certain experimental circumstances (see in Fig.~1 the
parts of the curves on the r.h.s. of the dashed lines). The value of
$\varphi_{\mbox{\scriptsize eq}}$ and the metastable region decrease
with the increase of the film thickness, which shows that the first
order of the phase transition is better pronounced in thinner films and
that confirms a conclusion in Ref.~\cite{Shopova2:2003}.

\begin{figure}
\begin{center}
\epsfig{file=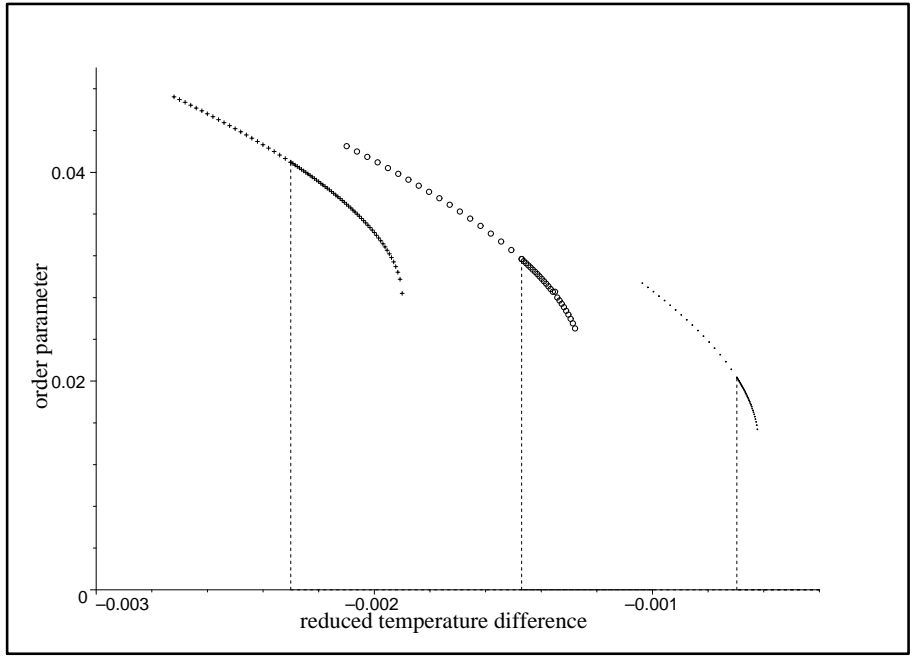,angle=-90, width=12cm}\\
\end{center}
\caption{Order parameter profile $\varphi(t_{0})$ of Al films of
different thicknesses: $L_0 = 0.05~\mu$m (``+"-line), $L_0 = 0.1~\mu$m
($\circ$), and $L_0 = 0.3~\mu$m ($\cdot$).} \label{ST2f1.fig}
\end{figure}

\begin{figure}
\begin{center}
\epsfig{file=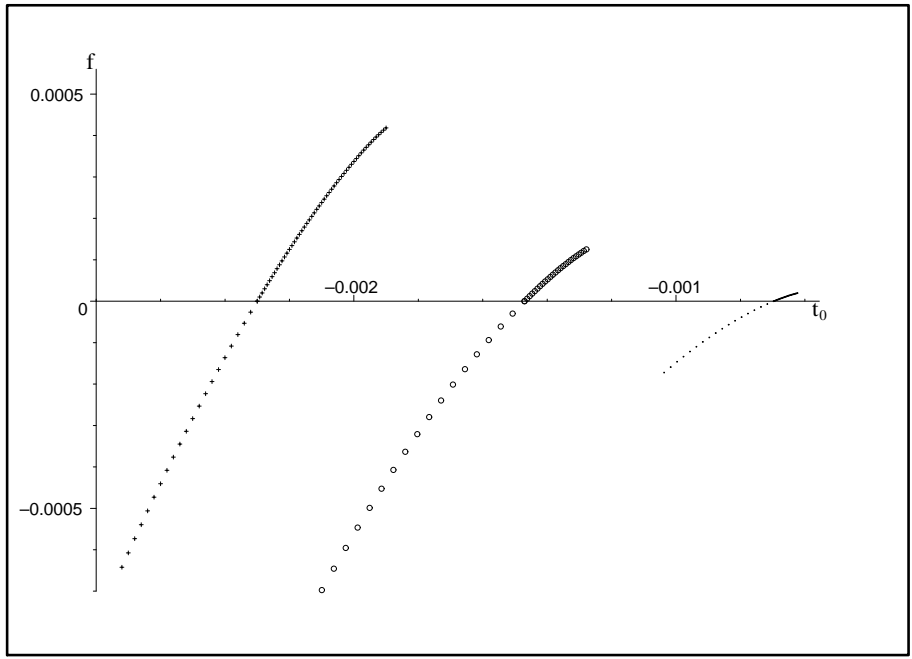,angle=-90, width=12cm}\\
\end{center}
\caption{The free energy $f(t_0)$ for Al films of thickness: $L_0 =
0.05~\mu$m (``+"-line), $L_0 = 0.1~\mu$m ($\circ$), $L_0 = 0.3~\mu$m
($\cdot$).}
 \label{ST2f2.fig}
\end{figure}

\begin{figure}
\begin{center}
\epsfig{file=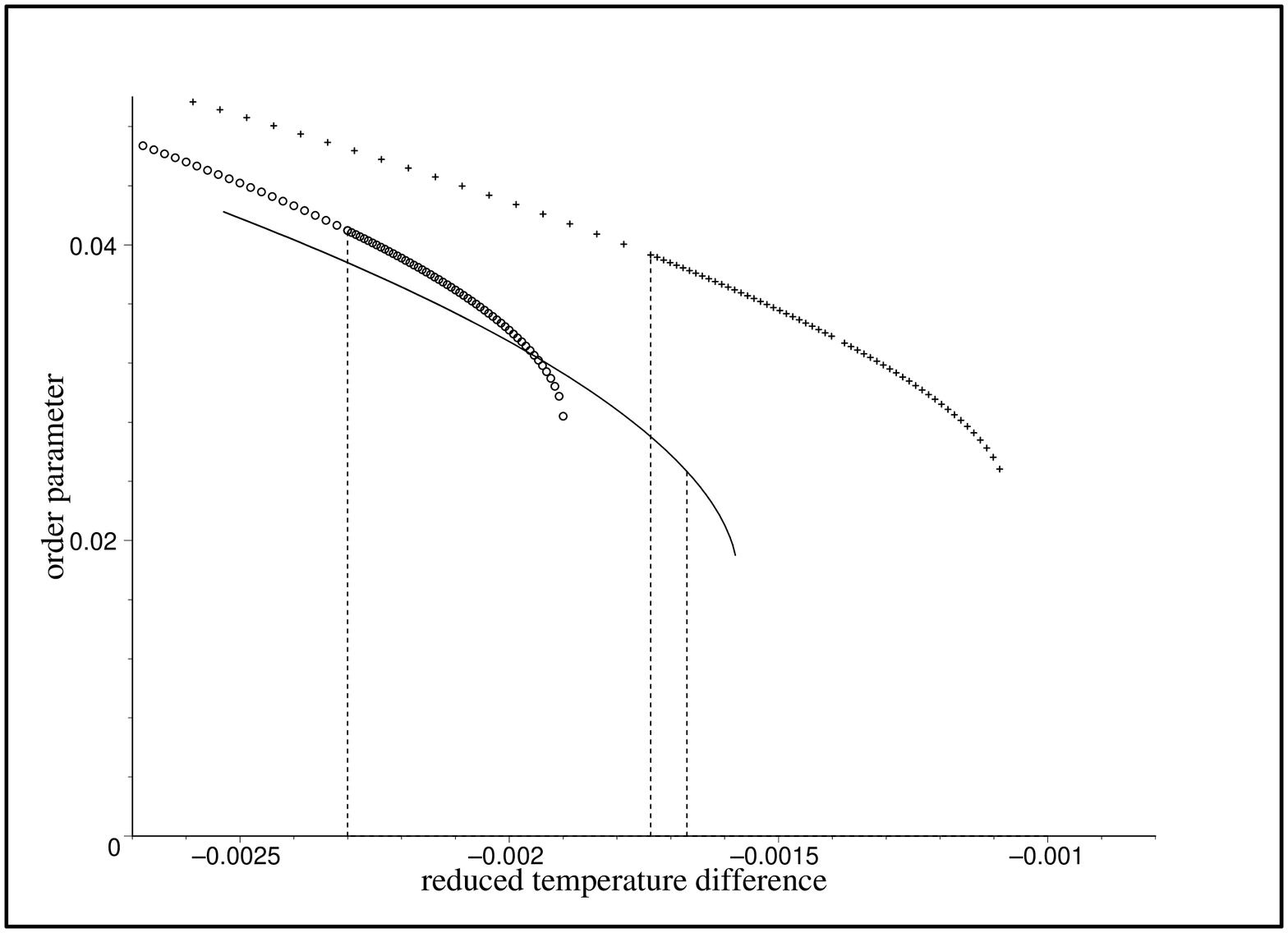,angle=-90, width=12cm}\\
\end{center}
\caption{Order parameter profile $\varphi(t_0)$ of films of thickness
$L_0 = 0.05~\mu$m: W (``+"-line), Al ($\circ$), and In ($\cdot$).}
\label{ST2f3.fig}
\end{figure}

These results are confirmed by the behavior of the free energy as a
function of $t_0$. We used Eqs.~(\ref{eq2}) and ~(\ref{eq4}) for the
calculation of the equilibrium free energy $f[\varphi_0(t_0)]$. The
free energy for Al films with different thicknesses is shown in Fig.~2.
The equilibrium points $T_{\mbox{\scriptsize eq}}$ of the phase
transition correspond to the intersection of the $f(\varphi_0)$-curves
with the $t_0$-axis. It is obvious from Fig.~2 that the temperature
domain of overheated Meissner states decreases with the increase of the
thickness $L_0$.

The shape of the equilibrium order parameter $\varphi_0$ in a broad
vicinity of the equilibrium phase transition of thin films ($L_0 =
0.05\mu$m) of W, Al, and In was found from Eq.~(\ref{eq4}). The result
is shown in Fig.~3.  The vertical dashed lines in Fig.~3 again indicate
the respective values of $t_{\mbox{\scriptsize eq}} =
t_0(T_{\mbox{\scriptsize eq}})$, at which the equilibrium phase
transition occurs as well as the equilibrium jump
$\varphi_0(T_{\mbox{\scriptsize eq}})= \varphi_{\mbox{\scriptsize eq}}$
in the different superconductors.

   The order parameter jump at the phase
transition point of In (the In curve is marked by points in Fig.~3) is
relatively smaller than for W, and Al, where the GL parameter has much
lower values. The same is valid for the metastability domains; see the
parts of the curves in Fig.~3 on the left of the vertical dashed lines.
It is obvious from Fig.~3 and Table~2 that the equilibrium jump of the
reduced order parameter $\varphi_{\mbox{\scriptsize eq}}$ of W has a
slightly smaller value than that of Al although the GL number $\kappa$
for W has a ten times lower value compared with $\kappa$ of Al. Note,
that in Fig~3 we show the jump of $\varphi_{\mbox{\scriptsize eq}}$,
but the important quantity is $|\psi|_{\mbox{\scriptsize eq}} =
|\psi_0|\varphi_{\mbox{\scriptsize eq}}$. Using the data  for $L_0 =
0.05\mu$m from Tables~1 and 2 we find for $|\psi|_{\mbox{\scriptsize
eq}}$ the following values: $0.1\times 10^{11}$ for Al, $0.05\times
10^{11}$ for In, and $0.02\times 10^{11}$ for W. This result shows that
the value of the critical filed $H_{c0}$ is also important and should
be taken into account together with the smallness of GL number when the
maximal values of the order parameter jump are looked for. Thus the
value of the order parameter jump at the fluctuation-induced phase
transition is maximal provided small values of the GL parameter
$\kappa$ are combined with relatively large values of the critical
field $H_{c0}$. In our case  Al has the optimal values of these two
parameters.

The shift of the phase transition temperature $t_{\mbox{\scriptsize
eq}} = |(T_{\mbox{\scriptsize eq}}-T_{c0})|/T_{c0}$, the reduced value
$\varphi_{\mbox{\scriptsize eq}}$ of the equilibrium order parameter
jump $|\psi|_{\mbox{\scriptsize eq}}$, and the latent heat $Q$ of the
equilibrium transition are given for films of different thicknesses and
substances in Table~2. The thicknesses are chosen so as to ensure the
validity of the theory~\cite{Folk:2001,Shopova2:2003} used in our
analysis and to satisfy other important requirements presented in
Sec.~4. The data in Table~2 show that the shift of the phase transition
temperature is very small and can be neglected in all calculations and
experiments based on them. The values for $\varphi_{\mbox{\scriptsize
eq}}$ for different $L_0$ and those for $|\psi_0|$ given in Table~1
confirm the conclusion which we have made for films of Al, In, and W
with $L_0 = 0.05\mu$m. The latent heat $Q$ has maximal values for In,
where the critical field is the highest for the considered materials.

{\bf 4.~Conclusion}

In contrast to our initial expectations that films made of
superconductors with extremely small GL parameter $\kappa$ such as Al
and, in particular, W will be the best candidates for an experimental
search of the HLM effect, our careful analysis  definitely gives
somewhat different answer. The Al films still remain a good candidate
for transport experiments through which the jump of the order parameter
at the phase transition point could be measured but surprisingly the W
films turn out inconvenient for the same reason because of  their very
low critical field $H_{c0}$. Although In has ten times higher GL number
$\kappa$ than Al, the In films can be used on an equal footing with the
Al films in experiments intended to prove the order parameter jump.
Here the choice of one of these materials may depend on other features
of experimental convenience. As far as caloric experiments are
concerned, the In films seem the best candidate for their high latent
heat.

We have presented the theoretical justification and predictions
intended to support experiments on the observation of magnetic
fluctuations and HLM effect near the normal-to superconducting
transition in a  zero external magnetic field. Besides, we have
demonstrated for the first time that the experiments can be most
successfully performed in type I superconductors with relatively high
critical magnetic fields.

{\bf Acknowledgements.} The authors thank Prof. D. I. Uzunov for useful
discussions. A funding by the European Superconductivity Network
(Scenet, Parma) and a research contract with JINR-Dubna are
acknowledged.

\end{document}